\input{psfig.sty}
\documentclass{article}
\begin{document}
\begin{center}
{\Large Spectral and temporal properties of X-ray emission from the
ultra-luminous source X-9 in
M81}
 \vspace{1cm} \\ 
{\bf V.~La Parola}, Universit\'a di Palermo; {\bf G.~Fabbiano}, Smithsonian CfA, 
Cambridge (MA); {\bf D.W.~Kim}, Smithsonian CfA, Cambridge (MA); {\bf G.~Peres},
Universit\'a di Palermo.\\
\end{center}

\begin{abstract}
We have analysed the spectra and the variability of individual 
X-ray sources in the M-81 field using data from the available ROSAT-PSPC and 
ROSAT-HRI observations of this nearby spiral galaxy. \\

Here we present the results on the second brightest source in the field 
(X-9 - Fabbiano, 1988 ApJ 325 544), whose identification and interpretation is 
still unclear. Our work includes the study of the shape of X-9 from HRI data,
the light curve and hardness ratio evolution, and the spectral analysis.
\end{abstract}
\section{Fundamental characteristics}
X9-M81, a very bright X-ray source, is located 12.5' far from M81 nucleus, at
coordinates  09h 57m 54s +69° 03' 48" (J2000). It was first detected with 
Einstein and, since then many hypotheses have been made on its nature, also on 
the basis of the very high X-ray luminosity (more than $10^{39}$ if at the 
distance of M~81). The most important ones are:
\begin{itemize}
\item      an extended structure (a super-shell) heated by the detonation of 
many ($>50$) supernovae (Miller 1995)
\item      a compact object (WD, NS or BH) accreting material from a dense 
cloud of interstellar gas (Miller 1995)
\item      a background quasar (Ishisaki et al. 1996)
\end{itemize}

Some important issues, that make understanding this object even more
difficult concern its optical appearance and its position. X-9 optical 
counterpart is not well identified: the only candidate so far is a weak 
(probably extended) object of $m_B\sim19$. Moreover, the source appears to be 
isolated,
well outside the M81 optical image and there is no evident association
with other galaxies in the group, albeit it may belong to the outskirts 
regions of the HoIX dwarf galaxy.
\newpage
\section{The data}
We analysed the available archival data from ROSAT PSPC (12 observation, with 8
pointed on M~81 nucleus for a total exposure time of 146 ksec) and HRI (7 
observations pointed on M~81 nucleus for a total exposure time of 135 ksec). 
We also used a SAX observation pointed on M~81 and one ASCA observation pointed 
on X-9.
\begin{table}[h]
\begin{center}
\begin{tabular}{|c|r|r|r|r|r|}\hline
instr.    &ref  &sequence nr&live time&start date&off axis\\ \hline\hline
ROSAT PSPC& 1P  &rp600101a00&    9296&25/03/91& 12.5'\\
          & 2P  &rp600110a00&   12717&27/03/91& 38.0'\\
          & 3P  &rp600052n00&    6588&18/04/91& 31.2'\\
          & 4P  &rp600110a01&   12238&15/10/91& 38.0'\\
          & 5P  &rp600101a01&   11085&16/10/91& 12.5'\\
          & 6P  &rp600382n00&   27120&29/09/92& 12.5'\\
          & 7P  &rp180015n00&   17938&03/04/93& 12.5'\\
          & 8P  &rp180015a01&    8731&04/05/93& 12.5'\\
          & 9P  &wp600576n00&   16412&29/09/93& 32.0'\\
          &10P  &rp180035n00&   17800&01/11/93& 12.5'\\
          &11P  &rp180035a01&    4234&07/11/93& 12.5'\\
          &12P  &rp180050n00&    1849&31/03/94& 12.5'\\ \hline
ROSAT HRI & 1H  &rh600247n00&   26320&23/10/92& 12.5'\\ 
          & 2H  &rh180015n00&    1688&16/04/93& 12.5'\\ 
          & 3H  &rh600247a01&   21071&17/04/93& 12.5'\\ 
          & 4H  &rh600739n00&   19902&19/10/94& 12.5'\\ 
          & 5H  &rh600740n00&   18984&13/04/95& 12.5'\\ 
          & 6H  &rh600881n00&   14826&12/10/95& 12.5'\\
          & 7H  &rh601002n00&   19776&30/09/97& 12.5'\\
          & 8H  &rh601095n00&   12590&25/03/98& 12.5'\\ \hline
BeppoSAX  &MECS &40732001   &  100287&04/06/98&  13' \\ 
          &LECS &           &   43931&        &      \\ \hline
ASCA      &SIS  &57048000   &   33000&06/04/99&  -   \\ \hline
\end{tabular}
\caption{Log of ROSAT, SAX, ASCA observation of X-9. Sequence 2H has not 
used for this work.}
\label{log} 
\end{center}
\end{table}

\newpage
\begin{figure}[h]
\caption{{\bf 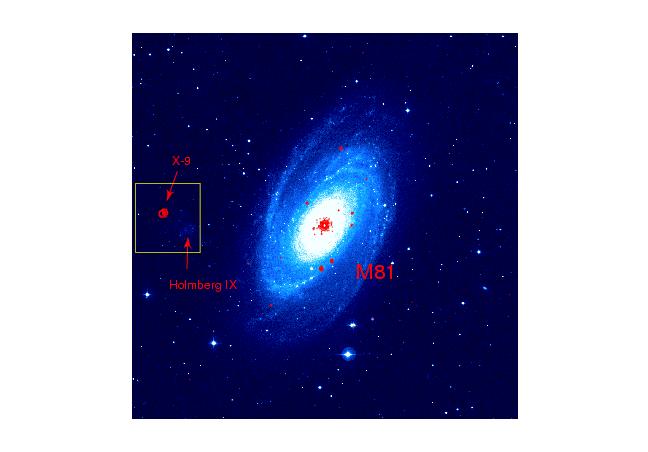 }: contour plots of X-ray count rates, superimposed to 
the optical image of M81 field. X-9 is marked by an arrow. 
{\bf  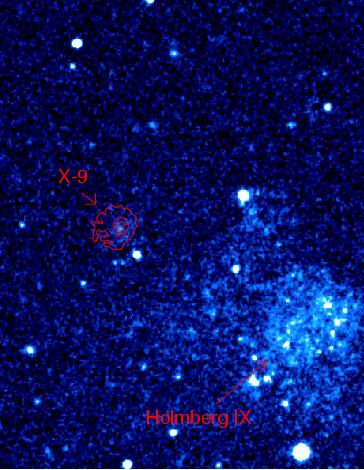}: 
Enlargement of the region within the box in the upper panel shows 
both the position of X-9
with respect to Holmberg IX and the faint optical source coincident with the
X-ray emission centroid}
\label{fov}
\end{figure}
\clearpage

\section{Spatial structure of the X-ray emission}
\begin{figure}[h]
\caption{{\bf fig2-up.jpg. Left}: contour plot of HZ43 counts in a ROSAT 
HRI image at 12 arcmin off axis. Note the asymmetric structure of the PSF.
{\bf Right}: contour plot of X-9 counts, rotated to have the same 
position as HZ43 with respect to the detector center. Sectors
used for the radial profile extraction are also shown. }
\label{calcont}
\centerline{\psfig{figure=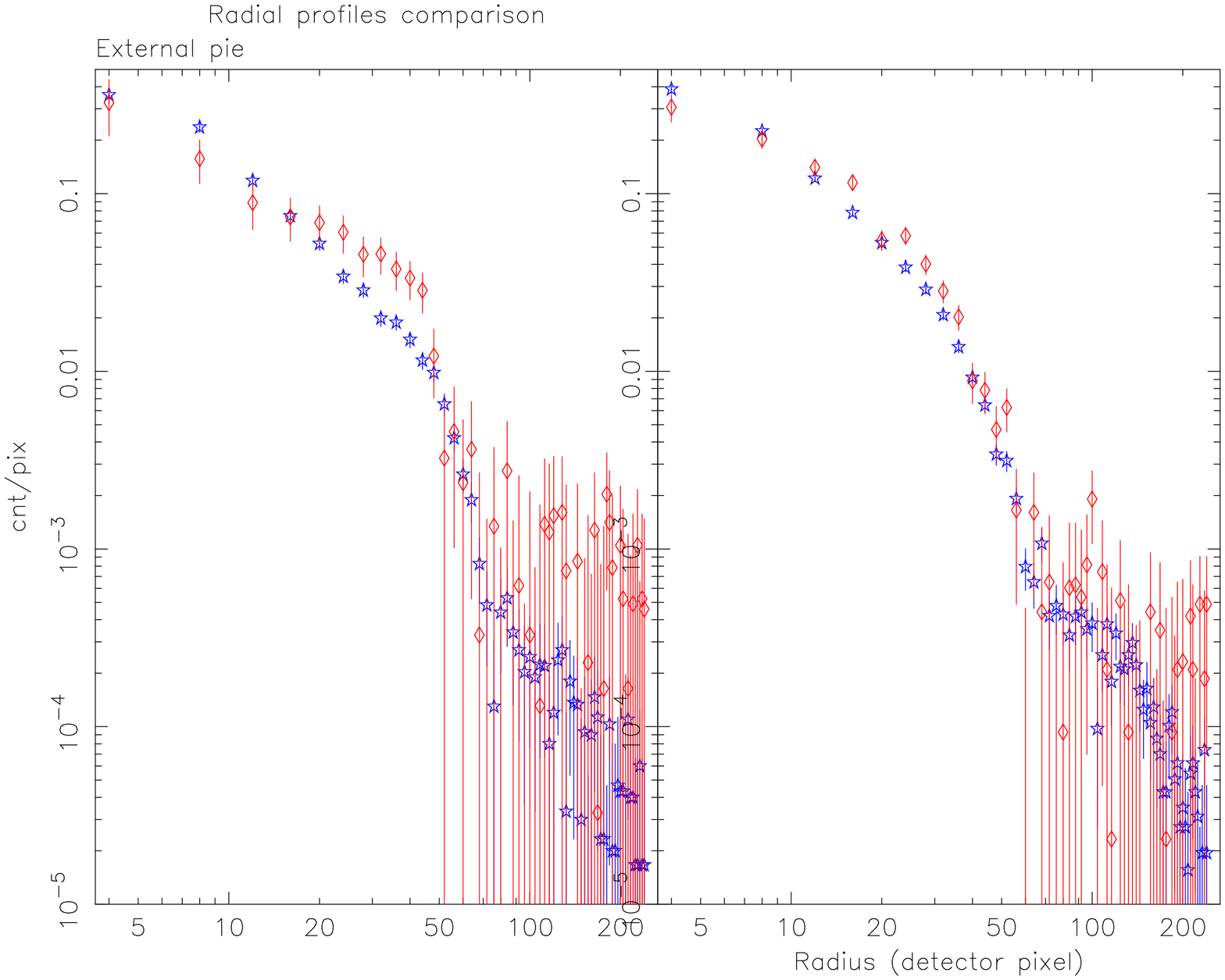,width=14cm,angle=270}}
\caption{Radial profiles of the calibration source HZ43 (stars), 
compared to X-9 profile (diamonds). Left panel: external sector. 
Right panel: internal sector. Data points are normalized to the total number 
of counts
within 2 arcmin. Extraction annuli are 2 arcsec wide (1 detector pixel = 0.5
arcsec)}
\label{rprof}
\end{figure}

The X-9 radial profile derived from HRI observations has been compared with 
that of an HRI
calibration source (HZ43) observed at the same off axis angle.\\
As the calibration profile is visibly asymmetric, we divided the source image 
into two pie slices (external/internal, see Figure~\ref{calcont}) with respect 
to the direction of the detector center. In the figure, X-9 has 
been rotated to have the same orientation has the calibration source. We 
derived the radial profile for each 
slice both for the calibration source and for X-9.  
Figure~\ref{rprof} shows the radial profile along the two directions. The right 
panel shows that X-9 
(diamonds) has a significant emission excess between 10 and 25 arcsec away 
from the center with respect to the calibration source (stars).\\
HRI data then suggest a very bright point-like source but also  
the presence of an extended component. 
\clearpage

\section{Source variability}
\begin{figure}[h]
\centerline{\psfig{figure=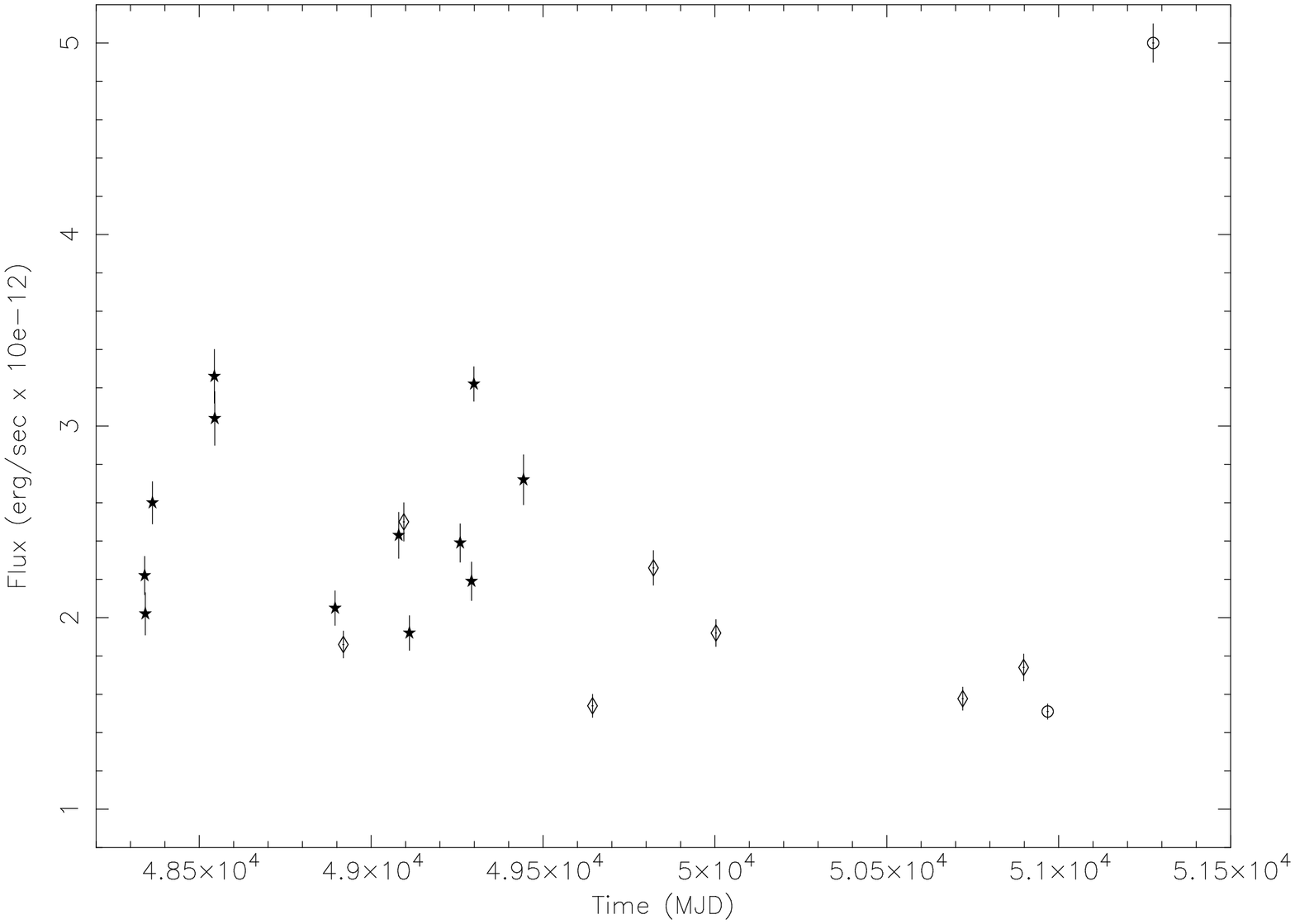,width=12cm,angle=270}}
\label{cntrate}
\caption{X-9 flux between 0.5 and 2.4 keV as a function of time. Stars are PSPC 
data points, diamonds
are HRI data points, the two circles represent the SAX (first in time) and the 
ASCA data points.}
\centerline{\psfig{figure=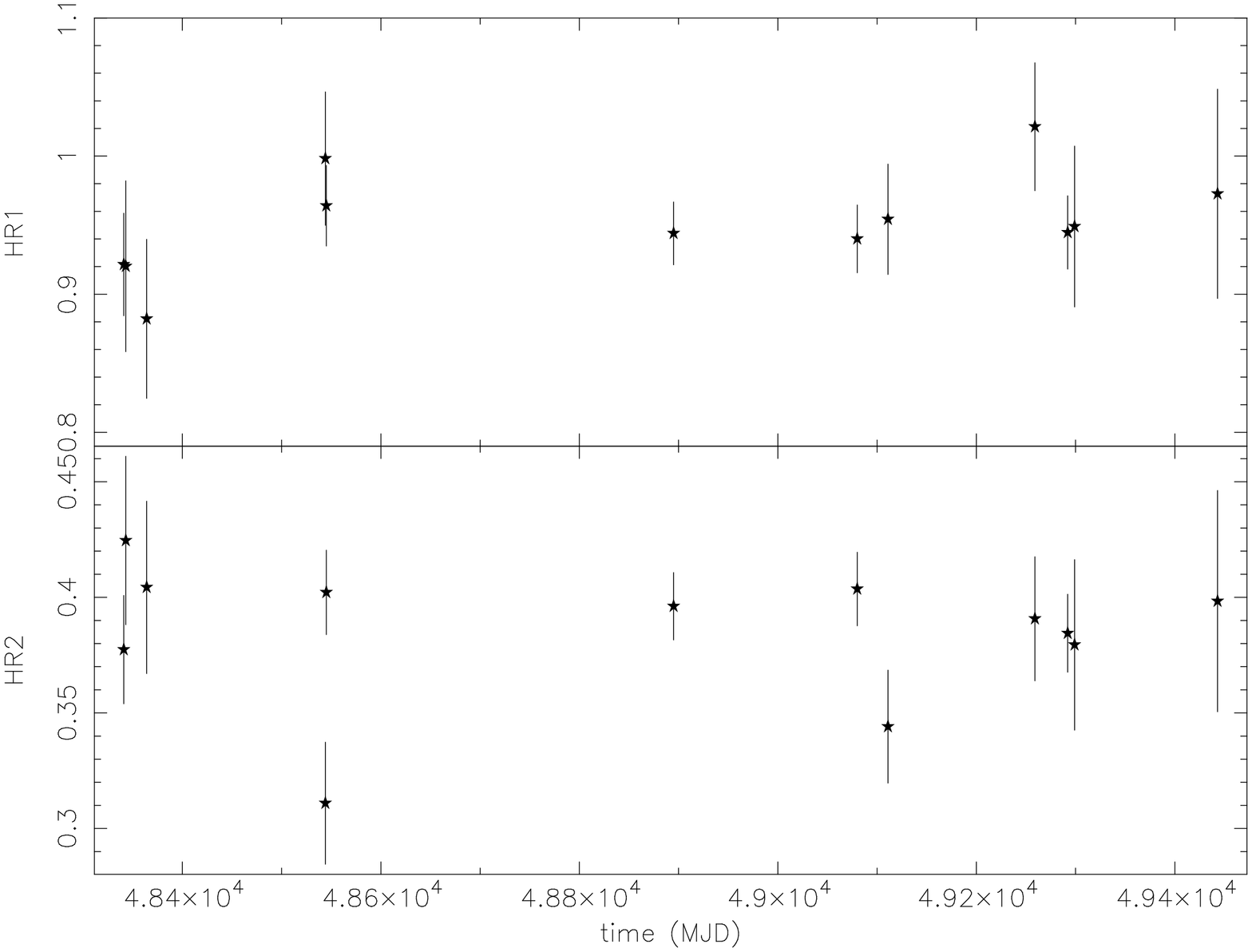,width=12cm,angle=270}}
\caption{Hardness ratios HR1 and HR2 for X-9 in M81 from ROSAT-PSPC data. 
Ratios HR1 and HR2 are both defined as $\frac{hard-soft}{hard+soft}$ for 
different definitions of the bands, see Table 2}
\label{hr}
\end{figure}

For each observation we calculated the source flux in the [0.5-2.4] keV energy 
band using the parameters that best fit the PSPC spectra (see next Section)
Results are reported in Figure~\ref{cntrate}, where each data point corresponds 
to one individual observation. Here we note that, in spite of the fact that they
are taken with different instruments, HRI and PSPC data are in
good agreement, and this is particularly evident in the period that covers 
April 1992. Moreover SAX data are consistent with the 
trend shown by the last HRI observations, while the ASCA point shows a much
higher emission. The source flux
is clearly variable: variability on time scales of months as well as a
descending trend on longer time scale are evident in Figure~\ref{cntrate}. The 
data do not 
show any hint of periodical variability. A flare or a change in the physical
status of the source could be responsible for the flux increase shown by  
the ASCA observation.\\
We also looked for spectral variability in PSPC data, by evaluating two 
different hardness ratios, both defined as 
\begin{center}
HR = $\frac{hard - soft}{hard + soft}$ 
\end{center}
using the energy bands showed in 
Table~\ref{hrdef}. We decided to use two different definitions of 
hardness ratio because the second (HR2) allows us a coarse spectral analysis 
of the high energy part of the ROSAT/PSPC band, appropriate for harder sources.
We plotted both hardness ratios in Figure~\ref{hr}, along with PSPC counts rate. 
We found very little evidence of variability, with only one point showing
a softer spectrum (observation 4P).\\
These results seem to exclude the hypothesis of a Supernovae shell: this
scenario could not explain the variability we observe unless we suppose a large
number of Supernovae explosions during the seven years covered by the observing 
period, but no 
event of this kind has been observed in other wavelenght bands. 
Instead, the light curve points to a compact source or an AGN.
\begin{table}[b]
\begin{center}
\begin{tabular}{|l l|c|c|} \hline
           &    &Soft            &Hard\\ \hline
Whole band &HR1 & 0.11-0.42 KeV  & 0.52-2.02 KeV\\ 
Hard band  &HR2 & 0.52-0.91 KeV  & 0.91-2.02 KeV\\ \hline
\end{tabular}
\caption{Definitions of bands used in the calculation of the two hardness 
ratios}
\end{center}
\label{hrdef}
\end{table}

\clearpage

\section{Spectral analysis}
We analysed spectral data derived from ROSAT/PSPC, SAX/LECS, SAX/MECS and
ASCA/SIS (GIS data were not used because of the contamination from M~81 nucleus)
In order to test the different hypotheses on the nature of X-9 we used the
following models:
\begin{itemize}
\item an absorbed power-law (implying the hypothesis of a quasar or of a 
compact object); 
\item a Raymond spectrum with one or two temperatures (related to the 
supershell hypothesis);
\item a multi-color black body disk (MCBB - adequate to a black hole 
hypothesis - Makishima et al., 2000). 
\end{itemize}
The Raymond model fails to fit the observed spectra.
The  best fit results are shown in Table~\ref{models}. Here we note 
that while SAX/LECS and ROSAT spectra appear very similar (in both cases the
MCBB best fit has been obtained using SAX/MECS to fix the hard power-law 
component), ASCA data show a considerably different behaviour, with a higher 
temperature disk and no evidence for the presence of a power-law or of a line. 
\begin{table}[h]
\begin{center}
\begin{tabular}{r|l|c|c|c|c|c}
&Instruments &nH($\times10^{22}$)  &$\Gamma$        & T (keV)
    &$E_{line}(\Delta E)$      &$\chi^2_{\nu}$ (Prob)\\ \hline\hline 
1&MECS        &$0.04^{+0.5} $       &$0.85\pm0.2         $ & - 
    &-                         & 1.11 (0.27)\\
2&MECS        &$0.04^{+0.6}      $  &$0.92\pm 0.16        $& -
    &$6.4^{+0.5}_{-0.17}(0.43)$& 1.03 (0.42)\\  \hline  

2&MECS+LECS   &$0.04^{+0.07}$     &$1.12^{+0.26}_{-0.14}$& -
    &$6.6^{+0.3}_{-0.8}(0.99)$& 1.04 (0.38)\\ 
3&MECS+LECS&$0.34^{+0.56}_{-0.26}$&$0.80^{+0.25}_{-0.40}$&$0.30^{+0.15}_{-0.12}$
    &$6.4^{+0.2}_{-0.6}(0.37)$& 0.91 (0.66)\\  \hline  

3&MECS+PSPC   &$0.16^{+0.13}_{-0.04}$&$0.82^{+0.27}_{-0.20}$&$0.32^{+0.07}_{-0.10}$ 
    &$6.4^{+0.25}_{-1.5}(0.39)$& 0.84 (0.83)\\   \hline 

4&ASCA SIS    &$0.177\pm0.015$       &-                     &$1.24\pm0.03$
    &-                         &1.13(0.05)\\
\end{tabular}
\caption{Spectral analysis: fit results. The first column contains the
model description according to the following code: 1) Power-law; 2)
Power-law + Gaussian line; 3) Power-law + MCBB + Gaussian line; 4) MCBB. 
The lower limit for the column
density is $4.1\times 10^{-2}$, i.e. the galactic line of sight value. $\Gamma$
is the power-law index. 
{\it T} is the temperature of the MCBB, wherever this model has been
used. The energy of the line is expressed in keV; we put within parenthesis the
line equivalent width in the same units. }
\label{models}
\end{center}
\end{table}
\begin{figure}[h]
\centerline{\psfig{figure=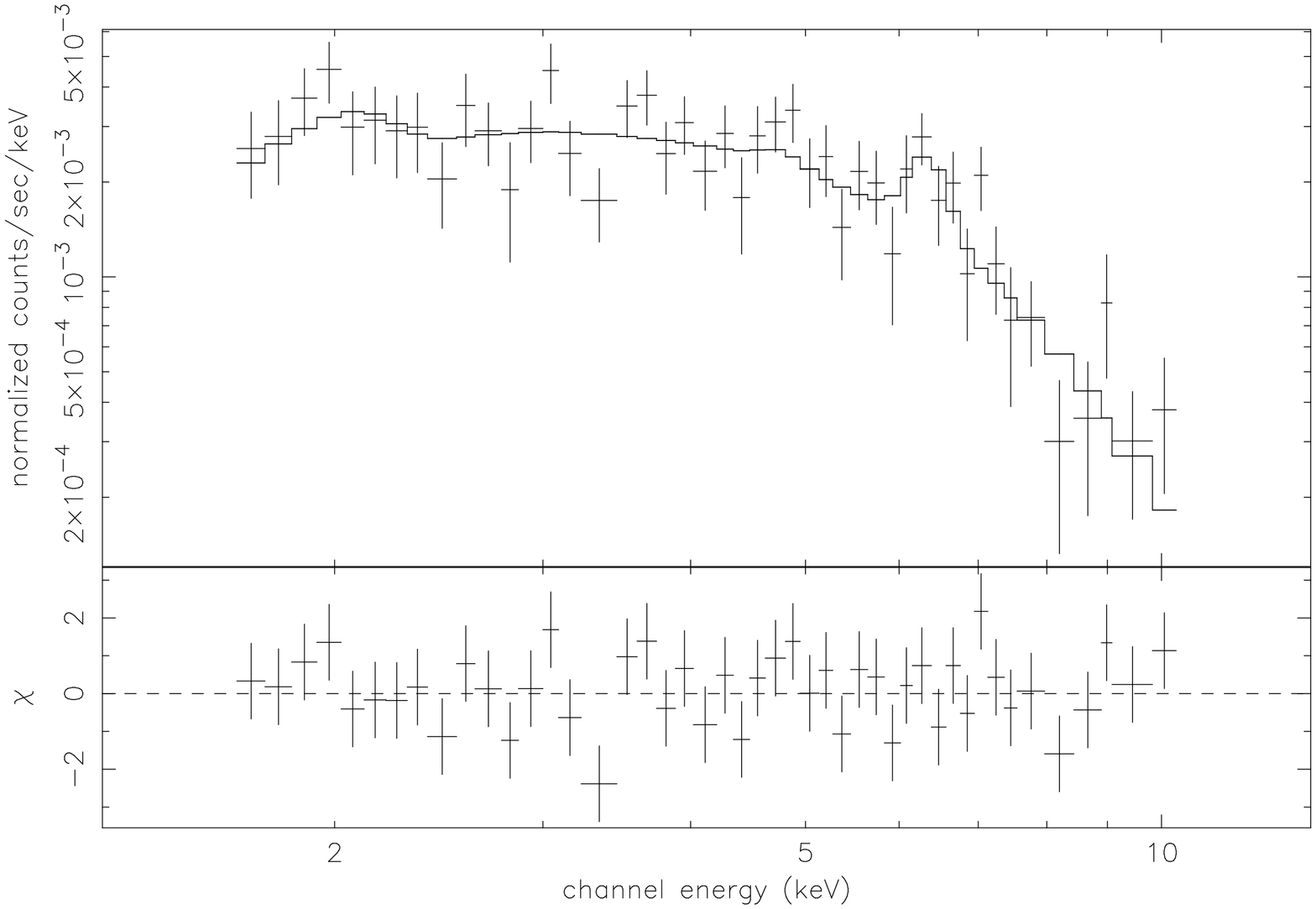,width=9cm,angle=270}
\psfig{figure=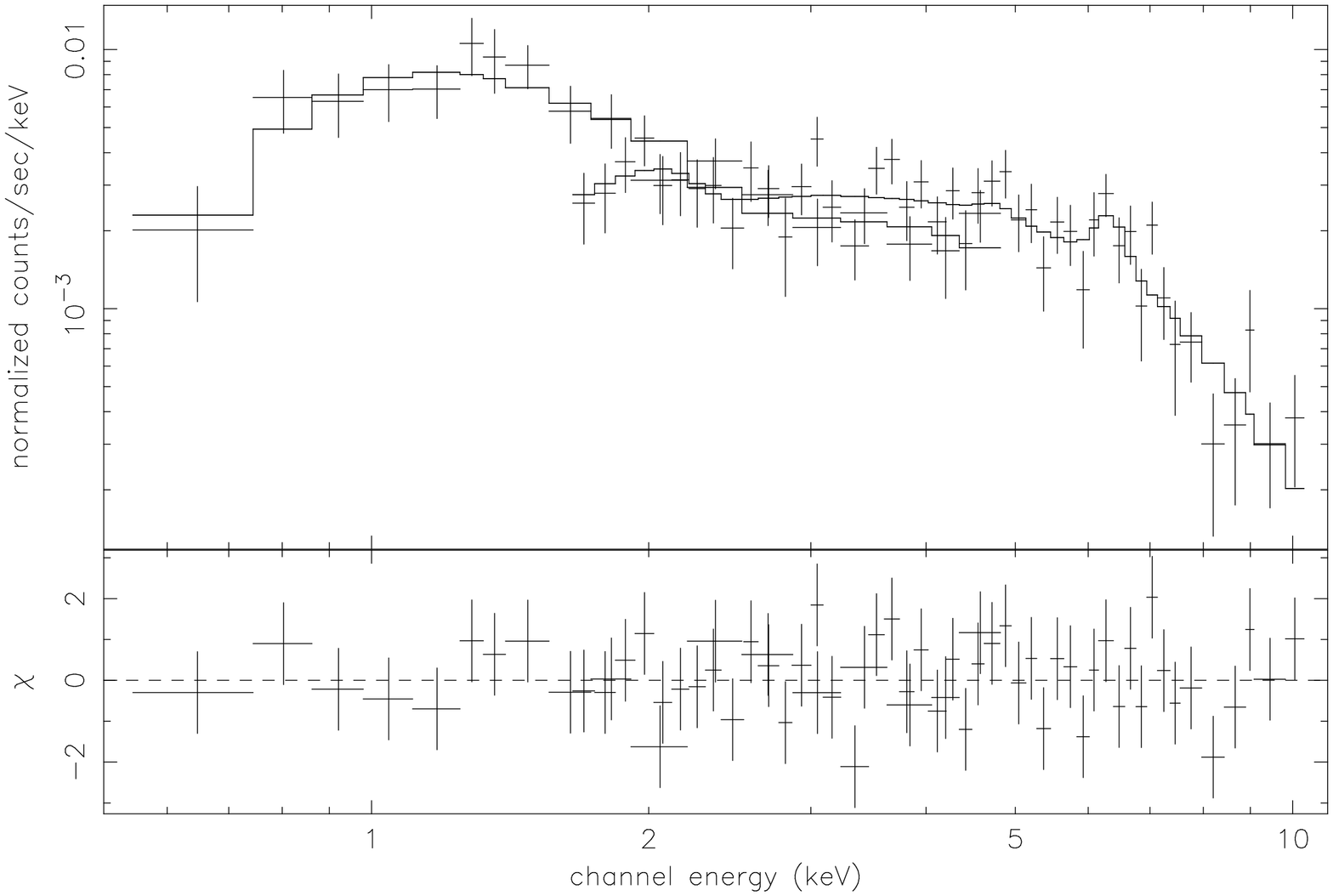,width=9cm,angle=270}}
\centerline{\psfig{figure=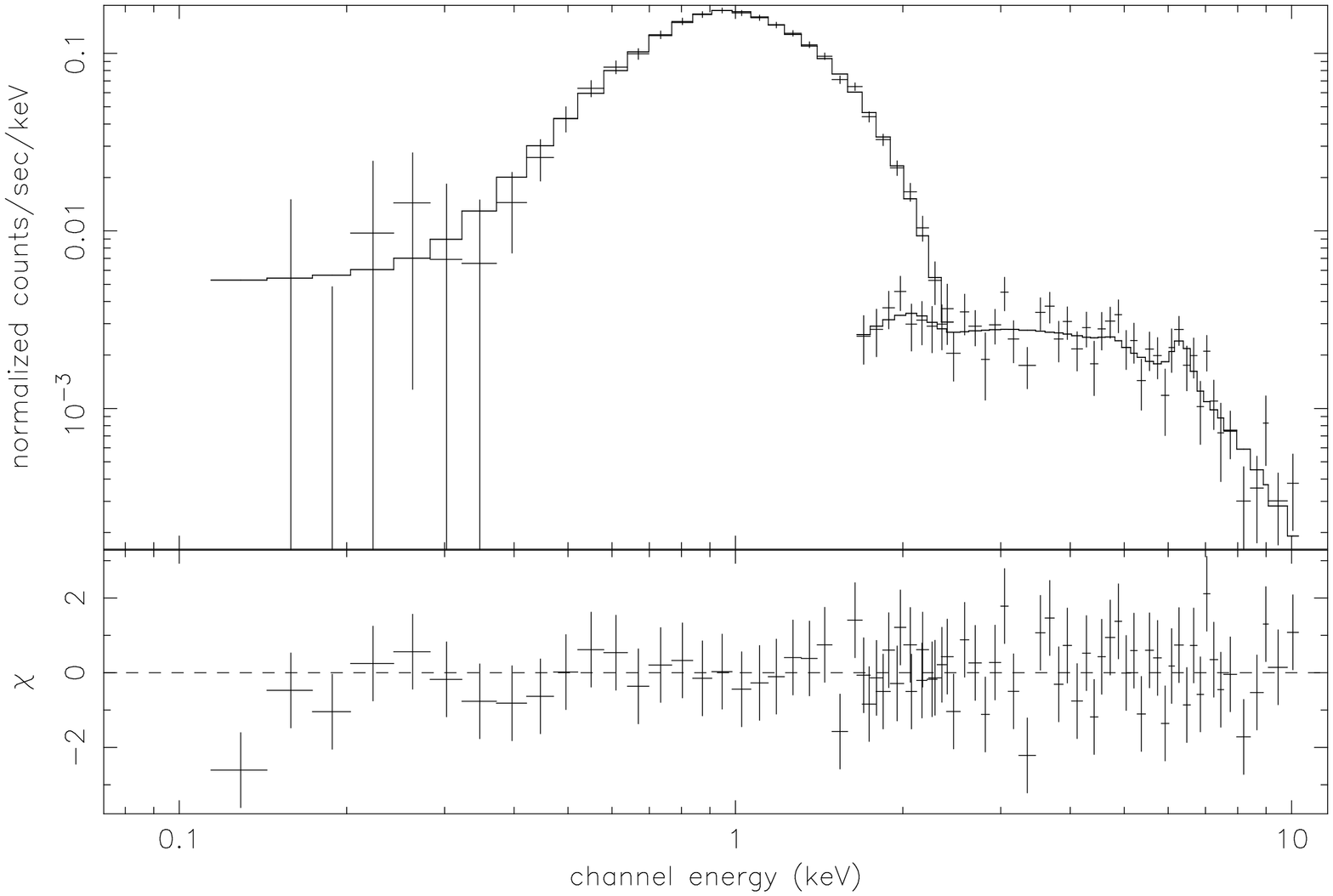,width=9cm,angle=270}
\psfig{figure=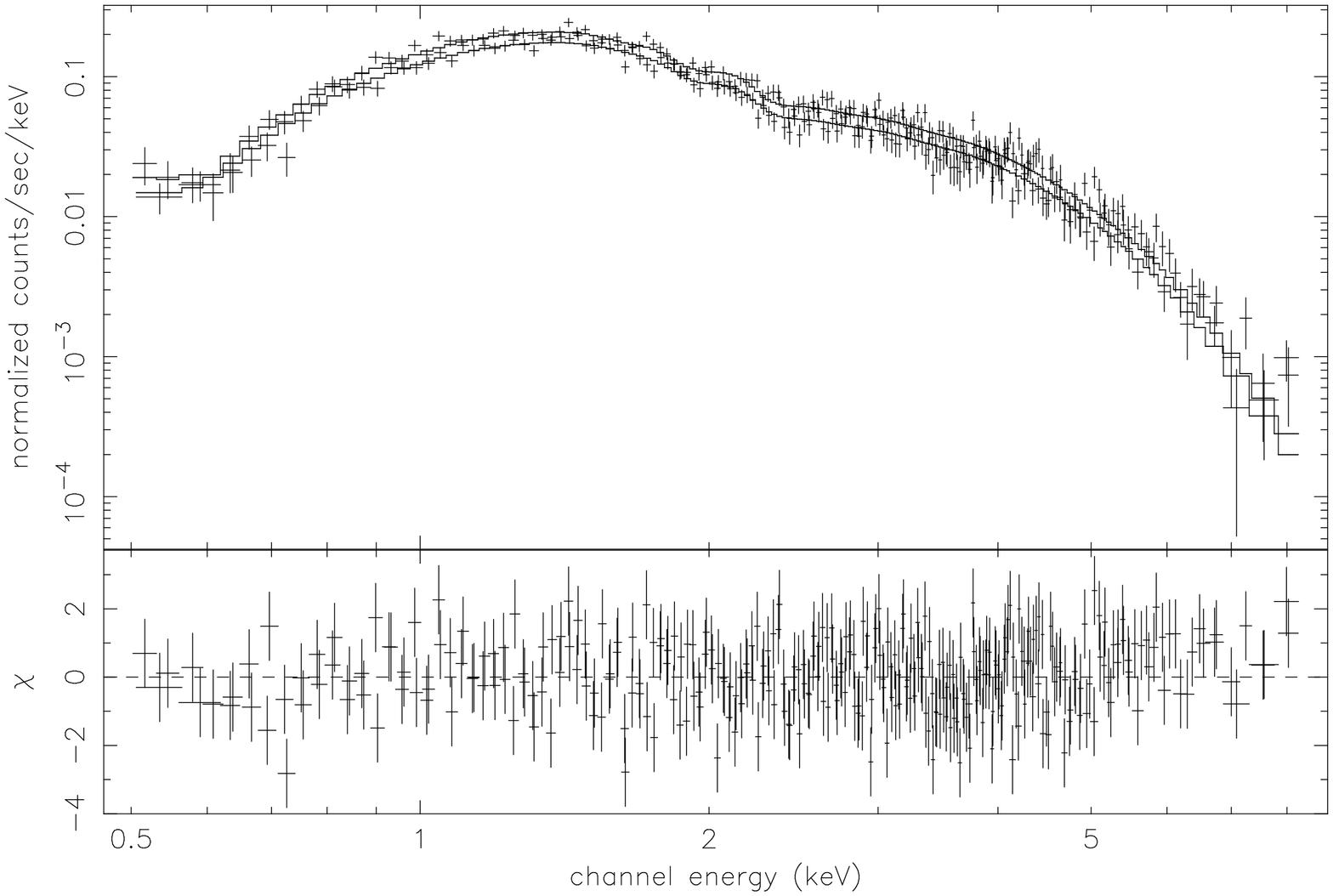,width=9cm,angle=270}}
\caption{X-9 spectra, best fit models and residuals in $\chi$ units. 
Upper left: MECS (power-law + line). Upper right: MECS + LECS (power-law + MCBB
+ line). Lower Left: PSPC + MECS (power-law + MCBB + line). Lower right: ASCA
(MCBB). The relevant parameters are in Table~\ref{models}. ASCA data are from the two
SIS detectors.}
\label{bestfit}
\end{figure}
\clearpage

\section{Final considerations}
\begin{itemize}
\item HRI emission centroid is well coincident with a faint object in the POSS 
II image
\item The source flux is strongly variable on relatively short time scales: 
this fact is 
at odds with the hypothesis of a Supernovae super shell and suggests that X-9 
is a compact accreting source
\item An extended component can however exist (as suggested by HRI data). 
\item If at M81 distance, the average X-ray luminosity of the source is $\sim 
4\times10^{39}$ erg/sec, well above the Eddington limit for a 1M$\odot$ 
accreting object
\item The results of the spectral fitting, supporting the MCBB model, seem to 
confirm the Super-Eddington nature of X-9: 
MCBB is the model used to fit the X-ray spectrum of Super-Eddington sources, 
and the temperature values we find are compatible with those of Galactic 
Black Hole candidates (Makishima et al. 1999)
\item A definite spectral variation is evident between SAX/PSPC and ASCA 
observations.
\end{itemize}
A more detailed description of the analysis, as well as the interpretation of
the results will appear in La Parola et al. (2000), in preparation.\\ 

This work was supported in
part by NASA grant NAG5-2946 and NASA contract NAS8-39073(CXC) and in part by
MURST. This research has made use of the HEASARC online database and of the ESO
online DSS.
\section{Reference}
Fabbiano G., 1988, ApJ, {\bf 325}, 544\\ 
Ishisaki Y. et al. 1996, PASJ, {\bf 48}, 237\\
La Parola V., Fabbiano G., Kim D.W., Peres G., Bocchino F., 2001, in preparation\\
Makishima K. et al. 2000, ApJ, {\bf 535}, 632\\
Miller B. W., 1995, ApJ, {\bf 446}, L75\\

\end{document}